\documentclass[11pt]{article}
\pdfoutput=1

\usepackage[nosort]{cite}
\usepackage[bulletsep]{collref}

\usepackage[british]{babel} 

\usepackage{epsfig,rotating}
\usepackage{amsmath,amssymb,amsthm}
\usepackage{amsfonts}
\usepackage{mathrsfs}
\usepackage{bbm}
\usepackage{bm}

\usepackage{color}
\usepackage[textwidth = 430 pt, textheight = 630 pt]{geometry}
\definecolor{MyDarkBlue}{rgb}{0.15,0.25,0.45}

\usepackage[linktocpage=true]{hyperref}
\hypersetup{
colorlinks=true,
citecolor=MyDarkBlue,
linkcolor=MyDarkBlue,
urlcolor=MyDarkBlue,
pdfauthor={Martin Wolf},
pdftitle={Contact Manifolds,  Contact Instantons, and Twistor Geometry},
pdfsubject={hep-th}
breaklinks=true
}

\let\fn\footnote
\renewcommand{\footnote}[1]{\linespread{1.1}\fn{#1}\linespread{1.29}}

\makeatletter\renewcommand{\section}{\@startsection
{section}{1}{\z@}{-3.5ex plus -1ex minus
    -.2ex}{2.3ex plus .2ex}{\bf\mathversion{bold} }}
\makeatletter\renewcommand{\subsection}{\@startsection{subsection}{2}{\z@}{-3.25ex
plus -1ex minus
   -.2ex}{1.5ex plus .2ex}{\bf\mathversion{bold} }}
\makeatletter\renewcommand{\subsubsection}{\@startsection{subsubsection}{3}{-2.45ex}{-3.25ex
plus -1ex minus -.2ex}{1.5ex plus .2ex}{\it }}
\renewcommand{\thesection}{\arabic{section}}
\renewcommand{\thesubsection}{\arabic{section}.\arabic{subsection}}
\renewcommand{\@seccntformat}[1]{\@nameuse{the#1}.~~}

\renewcommand{\theequation}{\thesection.\arabic{equation}}
\makeatletter \@addtoreset{equation}{section}

\renewcommand*\l@section{\@dottedtocline{1}{0em}{2em}}
\renewcommand*\l@subsection{\@dottedtocline{2}{2em}{2.4em}}
\renewcommand*\l@subsubsection{\@dottedtocline{4}{3.8em}{3.7em}}

\renewcommand\tableofcontents{%
    \section*{\large\contentsname
        \@mkboth{%
          \MakeUppercase\contentsname}{\MakeUppercase\contentsname}}%
       {\baselineskip=15pt plus 2pt minus 1pt
    \@starttoc{toc}}%
}

\renewenvironment{thebibliography}[1]
     {\baselineskip=16pt plus 2pt minus 1pt
      \section*{\large\refname
        \@mkboth{\MakeUppercase\refname}{\MakeUppercase\refname}}%
     \list{\@biblabel{\@arabic\c@enumiv}}%
           {\settowidth\labelwidth{\@biblabel{#1}}%
            \leftmargin\labelwidth
            \advance\leftmargin\labelsep
            \@openbib@code
            \usecounter{enumiv}%
            \let\p@enumiv\@empty
            \renewcommand\theenumiv{\@arabic\c@enumiv}}%
      \sloppy
      \clubpenalty4000
      \@clubpenalty \clubpenalty
      \widowpenalty4000%
      \sfcode`\.\@m
 \catcode`\^^M=10%
}

\newcommand{\appendices}{
\section*{Appendix}\label{appendices}\setcounter{subsection}{0}
\addcontentsline{toc}{section}{Appendix}
\setcounter{equation}{0}
\makeatletter
\renewcommand{\theequation}{\Alph{subsection}.\arabic{equation}}
\renewcommand{\thesubsection}{\Alph{subsection}}
\@addtoreset{equation}{subsection}
\makeatother
}

\numberwithin{lemma}{section}

\numberwithin{definition}{section}

\newtheorem{theorem}{Theorem}
\numberwithin{theorem}{section}

\numberwithin{prop}{section}

\numberwithin{cor}{section}

\newtheorem{remark}{Remark}
\numberwithin{remark}{section}

\def\periodb#1{\setbox0=\hbox{$#1$}#1\hskip-\wd0\hbox to\wd0{-}}

\newcommand{\CA}{\mathcal{A}}    			

\newcommand{\CC}{\mathcal{C}}

\newcommand{\CF}{\mathcal{F}}

\newcommand{\CH}{\mathcal{H}}

\newcommand{\CI}{\mathcal{I}}

\newcommand{\CN}{\mathcal{N}}

\newcommand{\CR}{\mathcal{R}}

\newcommand{\CT}{\mathcal{T}}

\newcommand{\CE}{\mathcal{E}}

\newcommand{\FR}{\mathbbm{R}}     			
\newcommand{\FC}{\mathbbm{C}}     			

\newcommand{\NN}{\mathbbm{N}}     			
\newcommand{\RZ}{\mathbbm{Z}}     			
\newcommand{\CPP}{{\mathbbm{C}P}}    			
\newcommand{\PP}{{\mathbbm{P}}}    			

\newcommand{\dd}{\mathrm{d}}     			
\newcommand{\dpar}{\partial}     			

\newcommand{\eand}{{~~~\mbox{and}~~~}}     		
\newcommand{\ewith}{{~~~\mbox{with}~~~}}
\newcommand{\efor}{{~~~\mbox{for}~~~}}

\newcommand{\der}[1]{\frac{\dpar}{\dpar #1}}   		

\newcommand{\sU}{\mathsf{U}}     			

\newcommand{\sSU}{\mathsf{SU}}

\newcommand{\sGL}{\mathsf{GL}}

\newcommand{\sSpin}{\mathsf{Spin}}

\def\tyng(#1){\hbox{\tiny$\yng(#1)$}}			
\def\tyoung(#1){\hbox{\tiny$\young(#1)$}}			

\begin{document}
\begin{titlepage}

\setcounter{page}{0}
\renewcommand{\thefootnote}{\fnsymbol{footnote}}

\begin{flushright}
DMUS--MP--12/03
\end{flushright}

\vspace{1cm}

\begin{center}

{\LARGE\textbf{\mathversion{bold}Contact Manifolds, Contact Instantons,  and Twistor Geometry}\par}

\vspace{1cm}

{\large
Martin Wolf
\footnote{{\it E-mail address:\/}
\href{mailto:m.wolf@surrey.ac.uk}{\ttfamily m.wolf@surrey.ac.uk}
}}

\vspace{1cm}

{\it 
Department of Mathematics\\
University of Surrey\\
Guildford GU2 7XH, United Kingdom

}

\vspace{1cm}

{\bf Abstract}
\end{center}
\vspace{-.3cm}

\begin{quote}
Recently, K\"all\'en \& Zabzine computed the partition function of a twisted supersymmetric Yang--Mills theory on the five-dimensional sphere using localisation techniques. Key to their construction is a five-dimensional generalisation of the instanton equation to which they refer as the contact instanton equation. Subject of this article is the twistor construction of this equation when formulated on $K$-contact manifolds and the discussion of its integrability properties. We also present certain extensions to higher dimensions and supersymmetric generalisations. 

\vfill
\noindent
30th June 2012

\end{quote}

\setcounter{footnote}{0}\renewcommand{\thefootnote}{\arabic{thefootnote}}

\end{titlepage}

\tableofcontents 

\bigskip
\bigskip
\hrule
\bigskip
\bigskip

\section{Introduction and results}

The self-dual Yang--Mills equation (or instanton equation) in four dimensions plays a very prominent role in both mathematics and physics. Over three decades ago, it was shown by Ward \cite{Ward:1977ta} (see also Atiyah \& Ward \cite{Atiyah:1977pw}) that all solutions to this equation on flat space-time have a natural interpretation in terms of holomorphic vector bundles over Penrose's twistor space  \cite{Penrose:1967wn,Penrose:1968me,Penrose:1969aa,Penrose:1972ia}. One often refers to this approach as the Penrose--Ward transform. Atiyah, Hitchin \& Singer generalised the Penrose--Ward transform to the curved setting \cite{Atiyah:1978wi} (see also \cite{Penrose:1976js,Ward:1980am}). In particular, they showed that the twistor space $Z\to M$ of an oriented Riemannian four-dimensional manifold $M$ comes equipped with a natural almost complex structure which is integrable if and only if the Weyl tensor of $M$ is self-dual. Solutions to the self-dual Yang--Mills equation on manifolds $M$ with self-dual Weyl tensor were then proven to be in one-to-one correspondence with holomorphic vector bundles over $Z$ that are holomorphically trivial up the fibres of $Z\to M$. For a detailed account on twistor theory for four-dimensional manifolds, we refer to the text books \cite{Manin:1988ds,Ward:1990vs,Mason:1991rf,Dunajski:2009aa}.

Recently, K\"all\'en \& Zabzine \cite{Kallen:2012cs} (see also Hosomichi, Seong \& Terashima\cite{Hosomichi:2012ek}) introduced a particular generalisation of the four-dimensional self-dual Yang--Mills equation that lives on five-dimensional contact metric  manifolds $M$.\footnote{Note that contact manifolds always admit contact metric structures  \cite{Blair:1976,Blair:2010}; see also Section \ref{sec:Kcontact}. When equipped with such a structure, they are called contact metric manifolds.} They refer to this generalised self-dual Yang--Mills equation as the {\it contact instanton equation}\footnote{To be more precise, we should distinguish between contact instantons and contact anti-instantons.}
\begin{equation}\label{eq:CIE}
  \CF\ =\ \pm\!\star_5\!(\eta\wedge\CF)\ =\ \pm\xi\lrcorner(\star_5\CF)~.
\end{equation}
Here, $\CF=\partial_\CA^2$ is the curvature for a connection $\partial_\CA$ represented locally by a connection one-form $\CA$ and $\eta$ is the contact form, $\xi$ the Reeb vector field, `$\star_5$' the Hodge star on $M$, and `$\lrcorner$' denotes contraction by a vector field. Concretely, K\"all\'en \& Zabzine studied the partition function of a twisted supersymmetric Yang--Mills theory on contact metric manifolds (in particular, on the five-dimensional sphere) using localisation techniques, and realised that the path integral of the five-dimensional theory localises on solutions to the contact instanton equation. It appears that in this supersymmetric setting, the Reeb vector fields needs to be Killing (which is the case for the five-dimensional sphere) \cite{Kallen:2012cs}. Contact metric manifolds whose Reeb vector field is Killing are called $K$-contact manifolds. Notice that in the special case when formulated on five-dimensional Sasaki--Einstein manifolds, the equation \eqref{eq:CIE} was already discussed by Harland \& N\"olle in  \cite{Harland:2011zs} (see also Tian \cite{Tian:2000}).

Subject of this article is to provide a detailed twistor construction of the contact instanton equation \eqref{eq:CIE} using ideas of Itoh's \cite{Itoh:2002,Itoh:2003}. We shall start by reviewing some basic properties of contact manifolds, contact metric manifolds and $K$-contact manifolds in Section \ref{sec:Kcontact}. In Section \ref{sec:TwistorSpace}, we then introduce Itoh's  Cauchy--Riemann twistor space $Z$ for five-dimensional $K$-contact manifolds $M$. This twistor space is a (real) seven-dimensional manifold that is fibred over $M$ with complex projective lines as typical fibres. It carries a natural almost Cauchy--Riemann structure whose integrability is determined by the vanishing of certain components of the curvature tensor on $M$ \cite{Itoh:2002,Itoh:2003}. This is rather similar to the above-mentioned case of four-dimensional Riemannian manifolds. In addition to this Cauchy--Riemann structure, we shall also be interested in a certain $F$-structure on $Z$ (in the sense of Rawnsley \cite{Rawnsley}) that is obtained by extending the Cauchy--Riemann structure by the horizontal lift of the Reeb vector field along the fibration $Z\to M$.  Throughout this article, we shall work in the spinor formalism. This is very natural from the twistor geometric point of view and as we shall see, this formalism allows us to present a very clear and short proof of Itoh's results, thereby making transparent all the geometric structures involved.  In Section \ref{sec:PWtrafo}, we then move on and establish Penrose--Ward transforms and Lax pairs including the one for the contact instanton equation \eqref{eq:CIE}.\footnote{The possibility of an extension of the Ward construction \cite{Ward:1977ta} was already mentioned in passing in \cite{Itoh:2002,Itoh:2003}.} The construction makes use of so-called partial connections that are induced by the afore-mentioned Cauchy--Riemann and $F$-structures.\footnote{$F$-structures are common to twistor constructions; e.g.~the twistor description of the three-dimensional (supersymmetric) Bogomolny monopole equation in terms of real geometries is based on $F$-structures \cite{Popov:2005uv}.}  
 Concretely, we shall see that  partially flat vector bundles over $Z$ that are holomorphically trivial up the fibres of $Z\to M$ are in one-to-one correspondence with solutions to the contact instanton equation (modulo gauge equivalences).

Generally, the contact instanton equation \eqref{eq:CIE} does not imply the Yang--Mills equation since the contact form is not closed. However, as shall be explained below, this equation admits essentially two different cases which seem different in nature: in the first case, \eqref{eq:CIE} is integrable by virtue of the twistor construction discussed in this article but it does not automatically  imply the Yang--Mills equation but rather the Yang--Mills equation with torsion.\footnote{Torsion Yang--Mills equations appear naturally in string theory, see e.g.~\cite{Grana:2005jc,Douglas:2006es,Blumenhagen:2006ci} and references therein. Our concrete example also appears when dimensionally reducing to five dimensions (anti-)self-dual three-form fields in six dimensions  \cite{Witten:2009at,Linander:2011jy}, see Remark \ref{rmk:action}.}  In the second case, \eqref{eq:CIE} does automatically imply the Yang--Mills equation but this case appears not to be accessible by the present twistor approach, and it remains to be seen if this case is integrable in general, as well.\footnote{If $M$ fibres over $S^1$ or $\FR$, then this case is also integrable. See below for details.}

In addition to five dimensions, we shall also discuss certain higher-dimensional generalisations and supersymmetric extensions of  \eqref{eq:CIE} using Bailey \& Eastwood's  idea of paraconformal geometries \cite{Bailey:1991xx}.

 We would like to point out that apart from the contact instanton equation \eqref{eq:CIE} there already exist many other higher-dimensional generalisations of the self-dual Yang--Mills equation in the literature,  see e.g.~\cite{Corrigan:1982th,Ward:1983zm,Donaldson:1985zz,Uhlenbeck:1986,Donaldson:1987,Capira:1988,Baulieu:1997jx,Donaldson:1998,Donaldson:2002,Popov:2010uq,Harland:2010ix,Popov:2010rf,Tian:2000}, including many solutions, see e.g.~\cite{Fairlie:1984mp,Corrigan:1984si,Fubini:1985jm,Ivanova:1992nj,Ivanova:1993ws,Gunaydin:1995ku} and more recently in e.g.~\cite{Popov:2009nx,Harland:2009yu,Correia:2009ri,Bauer:2010fia,Correia:2010xy,Haupt:2011mg,Gemmer:2011cp,Dunajski:2011sx,Ivanova:2012,Harland:2011zs}.
 
\section{Contact manifolds}\label{sec:Kcontact}

We start our discussion by reviewing some basics about $K$-contact manifolds. We shall be brief, however, and cover only material needed for our later discussion. For a detailed account on the subject, we refer to the text books by Blair \cite{Blair:1976,Blair:2010}  (see also the reviews by Boyer \& Galicki \cite{Boyer:2007nr} and Sparks \cite{Spark:2010}).

\paragraph{Contact manifolds.}
Let $M$ be a $(2m+1)$-dimensional smooth manifold. A contact structure on $M$ is a rank-$2m$ distribution $H\hookrightarrow TM$ in the tangent bundle of $M$. The distribution $H$ is called the contact distribution. The quotient of $TM$ by $ H$ yields a line bundle called the contact line bundle $L:=TM/ H$. Alternatively, the contact distribution can also be defined dually to be the kernel of a nowhere vanishing differential one-form $\eta$, called the contact form, which is defined up to scale on $M$, that is, $ H=\ker\eta$. If so, the contact form takes values in the line bundle $L$ since the map $TM\to L$ is the contraction of a vector field with $\eta$. We say that the contact structure is non-degenerate if for any two vector fields $X,Y\in\Gamma(M,H)$, the Frobenius form 
\begin{equation}  
 \Phi\,:\, H\wedge  H\ \to\ L\ =\ TM/ H~,\ewith \Phi(X,Y)\ :=\ [X,Y] \mbox{ mod }  H
\end{equation}
is non-degenerate on $H$. This is equivalent to saying that $\eta\wedge(\dd\eta)^m\neq0$ on all of $M$, where  $(\dd\eta)^m:=\dd\eta\wedge\cdots\wedge\dd\eta$ ($m$-times). Furthermore, there exists a unique vector field $\xi$, called the Reeb vector field, which obeys
\begin{equation}
 \xi\lrcorner\eta\ =\ 1\eand\xi\lrcorner\dd\eta\ = 0~.
 \end{equation}
Here, the symbol `$\lrcorner$' denotes contraction by vector fields. Sometimes, we shall also write $X\lrcorner\eta\equiv\eta(X)$. If $M$ is equipped with a non-degenerate contact structure, we call it a contact manifold. 

\paragraph{Contact metric manifolds.}
Next we wish to introduce the notion of a contact metric structure. Let $M$ be a $(2m+1)$-dimensional contact manifold with contact form $\eta$ and Reeb vector field $\xi$. Furthermore, let $g$ be a Riemannian metric on $M$. Then $(\eta,\xi,g,\phi)$ is called a contact metric structure and $M$ a contact metric manifold if $\phi$ is an endomorphism of $TM$ such that for any two vector fields $X,Y\in\Gamma(M,TM)$, we have
\begin{equation}\label{eq:CMS}
\begin{aligned}
 \phi^2(X)\ =\ -X+\eta(X)\xi~,\qquad
 g(X,\phi(Y))\ =\ \dd\eta(X,Y)~,\\
 g(\phi(X),\phi(Y))\ =\ g(X,Y)-\eta(X)\eta(Y)~.\kern1.2cm
 \end{aligned}
\end{equation}
This then implies that $\phi(\xi)=0$, $g(\xi,\xi)=1$, and $\eta(X)=g(X,\xi)$. We would like to point out that contact manifolds always admit contact metric structures  \cite{Blair:1976,Blair:2010}. This makes clear that the structure group of a contact manifold can be reduced to $\sU(m)\times 1$. 

\paragraph{Connection and curvature.}
After having talked about contact and contact metric structures, we move on and discuss connections and curvature. Let $\nabla$ be the Levi-Civita connection on a contact metric manifold $M$ and $\mbox{pr}:TM\to H$ the projection mapping. Then the contact distribution inherits a natural metric connection $\nabla^H$ given by\begin{equation}\label{eq:nablaH}
 \nabla_X^H Y\ :=\ \mbox{pr}(\nabla_XY)\efor X,Y\ \in\ \Gamma(M,TM)~.
\end{equation}
We have
\begin{equation}
\nabla_XY- \nabla^H_XY\ =\ -g(Y,\nabla_X\xi)\xi\efor X,Y\ \in\ \Gamma(M,TM)~,
\end{equation}
where $\xi$ is the Reeb vector field as before and $g$ the metric. This expression is not symmetric in $X$ and $Y$ since the connection $\nabla^H$ has torsion (recall that the contact distribution is maximally non-integrable): 
\begin{equation}\label{eq:TorsionH}
 \CT^H(X,Y)\ :=\ \nabla^H_XY-\nabla^H_YX-[X,Y]\ =\ -\eta([X,Y])\xi\efor X,Y\ \in\ \Gamma(M,TM)~.
\end{equation}
In addition, if we set
\begin{equation}\label{eq:curvatures}
\begin{aligned}
  \CR(X,Y)Z\ &:=\ \nabla_X\nabla_YZ -\nabla_Y\nabla_X Z-\nabla_{[X,Y]}Z~,\\
  \CR^H(X,Y)Z\ &:=\ \nabla^H_X\nabla^H_YZ -\nabla^H_Y\nabla^H_X Z-\nabla^H_{[X,Y]}Z
  \end{aligned}
\end{equation}
for $X,Y,Z\in\Gamma(M,TM)$, then a short calculation reveals that the Gau{\ss} equation is given by
\begin{eqnarray}\label{eq:Gauss}
 \CR^H(X,Y)Z\ &\!=\!&\ \CR(X,Y)Z-g(\CR(X,Y)Z,\xi)\xi -g(Z,\nabla_X\xi)\nabla_Y\xi+g(Z,\nabla_Y\xi)\nabla_X\xi  \notag\\
    &\!=\!&\ \mbox{pr}(\CR(X,Y)Z)-g(Z,\nabla_X\xi)\nabla_Y\xi+g(Z,\nabla_Y\xi)\nabla_X\xi~.
 \end{eqnarray}
To arrive at this expression, we have used that $g(\xi,\nabla_X\xi)=0$.

\paragraph{\mathversion{bold}$K$-contact manifolds.}
Next we would like to introduce the concept of $K$-contact manifolds. They form a special class of contact metric manifolds, that is, a contact metric manifold $M$ is called $K$-contact if the associated Reeb vector field is Killing (hence, the prefix `$K$'). 

If $M$ is $(2m+1)$-dimensional, we then have
\begin{equation}\label{eq:Kcontact1}
 \nabla_X\xi\ =\ -\phi(X)\efor X\ \in\ \Gamma(M,TM)
\end{equation}
and the curvature relations 
\begin{equation}\label{eq:Kcontact2}
 \CR(\xi,X)\xi\ =\ -X\efor X\ \in\ \Gamma(M, E)\eand\mbox{Ric}(\xi,\xi)\ =\ 2m~.
\end{equation}
The first of these relations follows from \eqref{eq:CMS} and the fact that $\xi$ is Killing,
\begin{eqnarray}
 g(X,\phi(Y))\! &=&\! \dd\eta(X,Y)\notag\\
   &=&\! \tfrac12\big((\nabla_X\eta)(Y)-(\nabla_Y\eta)(X)\big)\notag\\
   &=&\! \tfrac12\big(g(\nabla_X\xi,Y)-g(\nabla_Y\xi,X)\big)
   \ =\  -g(X,\nabla_Y\xi)
\end{eqnarray}
for $X,Y\in\Gamma(M,TM)$. The first equation of \eqref{eq:Kcontact2} is a consequence of \eqref{eq:Kcontact1} and $\phi^2(X)=-X$ for $X\in\Gamma(M, H)$ since
\begin{eqnarray}
 -X\ =\  \phi^2(X)\! &=&\! \nabla_{\nabla_X\xi}\xi\notag\\
  &=&\! \nabla_\xi\nabla_X\xi+[\nabla_X\xi,\xi]\ =\ \nabla_\xi\nabla_X\xi-\nabla_X\nabla_\xi\xi+[\nabla_X\xi,\xi]\notag\\
  &=&\! \CR(\xi,X)\xi
\end{eqnarray}
while the condition on the Ricci tensor is a direct corollary of this expression.

In particular, we see that the Ricci tensor of the Levi-Civita connection for $K$-contact manifolds in the direction of the Reeb vector field is constant and determined by the dimension of $M$. The converse is also true \cite{Blair:1976,Blair:2010}:

\vspace{15pt}
{\theorem 
A contact metric manifold $M$ of dimension $2m+1$ is K-contact if and only if the Ricci tensor in the direction of the Reeb vector field is equal to $2m$, that is, $\operatorname{Ric}(\xi,\xi)=2m$.
}
\vspace{15pt}

\paragraph{Examples.}
Examples of manifolds which are $K$-contact are plentiful. Perhaps the prime example is the so-called Boothby--Wang fibration when a contact manifold $(M,\eta,\xi)$ arises as a circle fibration over a symplectic manifold $N$. Briefly, the symplectic form on $N$, which needs to lie in $H^2(N,\RZ)$, pulls back to $\dd\eta$ on $M$, the endomorphism $\phi$ is induced by the almost complex structure on $N$ (that makes $N$ almost K\"ahler) while the Reeb vector field $\xi$ is Killing with respect to the metric that arises as the pull back of the metric on $N$ plus a term of the form $\eta\otimes\eta$. This metric then satisfies all of the conditions given in \eqref{eq:CMS}. A special case of a Boothby--Wang fibration is the Hopf fibration where odd-dimensional spheres arise as circle fibrations over complex projective spaces. Another very important class of $K$-contact manifolds is given by the class of Sasakian geometries: a $K$-contact manifold is called Sasakian, if 
\begin{equation}
 (\nabla_X\phi)(Y)\ =\ g(X,Y)\xi-g(\xi,Y)X~\efor X,Y\ \in\ \Gamma(M,TM)~.
\end{equation}
This is equivalent to saying that the curvature obeys
\begin{equation}\label{eq:CurvSas}
 \CR(X,\xi)Y\ =\ -g(X,Y)\xi+g(\xi,Y)X~\efor X,Y\ \in\ \Gamma(M,TM)~.
\end{equation}

\section{Twistor construction of contact manifolds}\label{sec:TwistorSpace}

Subject of this section is the construction of a twistor space for $K$-contact manifolds. To this end, we first restrict our focus to the five-dimensional setting and discuss Itoh's Cauchy--Riemann twistor space \cite{Itoh:2002,Itoh:2003}.  Using the formalism of (commuting) spinors, which is very natural from the twistor theoretic point of view, we are able to present very  concise and short proofs of his results.  At the end of this section, we shall extend this construction and discuss higher-dimensional generalisations  by making use of Bailey \& Eastwood's  idea of paraconformal geometries \cite{Bailey:1991xx}.

\paragraph{Conformal structures.}
Recall that when $M$ is a four-dimensional Riemannian spin manifold, then the complexified tangent bundle $T_\FC M:=TM\otimes\FC$ can be factorised into two rank-$2$ complex vector bundles $S$ and $\tilde S$, $\sigma:T_\FC M\overset{\cong}{\to} S\otimes\tilde S$. The bundles $S$ and $\tilde S$ are simply the bundles of anti-chiral and chiral spinors. These bundles come equipped with real structures $\tau:S\to S$ and $\tilde\tau:\tilde S\to\tilde S$ which are anti-linear and obey $\tau^2=-1=\tilde\tau^2$ such that i) the real structure $\tau_M$ induced on $T_\FC M:=TM\otimes\FC$ is given by $\tau_M=\tau\otimes\tilde\tau$ and obeys $\tau_M^2=1$ and ii) the set of fixed points of $\tau_M$ coincides with $TM$. Furthermore, the factorisation $\sigma:T_\FC M\overset{\cong}{\to} S\otimes\tilde S$ is equivalent to choosing a conformal structure on $M$, since this isomorphism yields naturally a complex line subbundle $\Lambda^2S\otimes\Lambda^2\tilde S$ in $T^\vee_\FC M\odot T^\vee_\FC M$.\footnote{We use `$\vee$' to denote the dual.}  The real structures on $S$ and $\tilde S$ extend to a real structure on $\Lambda^2 S\otimes\Lambda^2\tilde S$ which has fixed points. The set of fixed points defines a real line bundle which in turn gives the conformal structure. The twistor space of $M$ is then defined to be the projectivisation of one of these spinor bundles. It is worth noting that even when a factorisation of the tangent bundle into spinor bundles does not exist globally, the projectivisation of these bundles does and hence is well defined even in the case when $M$ is not spin.

In the present context of a five-dimensional contact manifold $M$, we wish to change this point of view slightly and introduce a conformal structure only on the contact distribution $H$. Therefore, we shall assume a factorisation of the form $\sigma:H_\FC\overset{\cong}{\to} S\otimes\tilde S$ into two rank-$2$ complex vector bundles $S$ and $\tilde S$ together with real structures $\tau$ and $\tilde\tau$ of the above type. We shall refer to this type of structure as a conformal contact structure and to contact manifolds equipped with such a structure as conformal contact manifolds.

Then we have the decomposition
\begin{equation}\label{eq:Decompo}
 \Lambda^2 H_\FC\ \cong\ \underbrace{\odot^2 S\otimes\Lambda^2\tilde S}_{=:\ \Lambda_+^2 H_\FC}\,\oplus\,\underbrace{\Lambda^2 S\otimes\odot^2\tilde S}_{=:\ \Lambda_-^2 H_\FC}~.
\end{equation}
The real structures $\tau,\tilde\tau$ extend to real structures on $\Lambda^2 H_\FC$ and $\Lambda^2_\pm H_\FC$ which, in turn, have fixed points on these bundles. We shall denote the corresponding set of fixed points by $\Lambda^2 H$ and $\Lambda^2_\pm H$, respectively. Correspondingly, we have a decomposition of differential two-forms $\Omega^2(M):=\Gamma(M,\Lambda^2 T^\vee M)$ with $\Omega^k_H(M):=\Gamma(M,\Lambda^k H^\vee)$ and $\Omega^2_\pm(M):=\Gamma(M,\Lambda^2_\pm H^\vee)$ according to
\begin{equation}\label{eq:TwoForms}
  \Omega^2(M)\ \cong\ \Omega^2_+(M)\oplus \Omega^2_-(M)\oplus \eta\wedge\Omega^1_H(M)~.
\end{equation}
 Notice that for any $k$-form $\omega\in\Omega^k(M)$, the part $\omega_H$ of $\omega$ lying in $\Omega^k_H(M)$ is given by $\omega_H=\omega-\eta\wedge(\xi\lrcorner\omega)$ while the $\omega_\eta$ part lying in $\eta\wedge\Omega^{k-1}_H(M)$ is $\omega_\eta=\eta\wedge(\xi\lrcorner\omega)$.  In the following, we shall write $\omega=\omega_++\omega_-+\omega_\eta$ for any two-form $\omega$ on $M$ with $\omega_\pm\in\Omega^2_\pm(M)$ and $\omega_\eta\in\eta\wedge\Omega^1_H(M)$. Notice that the two-forms $\Omega_\pm^2(M)$ simply represent the self-dual and anti-self-dual two-forms on $H$. 
   
We call the structure $(\eta,\xi,g,\phi)$ a conformal contact metric structure and therefore $M$ a conformal contact metric manifold if the relations \eqref{eq:CMS} are satisfied and $(\eta,\xi,g,\phi)$ is compatible with the conformal structure, that is, $\dd\eta$ sits in\footnote{From \eqref{eq:CMS} we have in particular that $\dd\eta(X,Y)=g(X,\phi(Y))$ and since $\phi|_H$ is an almost complex structure on $H$, $\dd\eta$ is either a self-dual or an anti-self-dual two-form on $H$. Assuming $\dd\eta\in\Omega^2_-(M)$ instead of $\dd\eta\in\Omega^2_+(M)$ merely corresponds to a change of orientation of $M$. Therefore, the assumption $\dd\eta\in\Omega^2_+(M)$ is no restriction.} $\Omega^2_+(M)$ and the metric $g$ restricts on $H$ to $g|_H=\varepsilon\otimes\tilde\varepsilon$, where $\varepsilon\in\Gamma_\tau(M,\Lambda^2 S^\vee):=\{\varepsilon\in \Gamma(M,\Lambda^2 S^\vee)\,|\,\tau(\varepsilon)=\varepsilon\}$ and $\tilde\varepsilon\in \Gamma_{\tilde\tau}(M,\Lambda^2\tilde S^\vee):=\{\tilde\varepsilon\in \Gamma(M,\Lambda^2\tilde S^\vee)\,|\,\tilde\tau(\tilde\varepsilon)=\tilde\varepsilon\}$.  In particular, we may choose frame fields $\{E_{\alpha\dot\alpha},\xi\}$ on $M$, with $\alpha,\beta,\ldots,\dot\alpha,\dot\beta,\ldots=1,2$ and $\xi$ the Reeb vector field, such that $\eta(E_{\alpha\dot\alpha})=g(E_{\alpha\dot\alpha},\xi)=0$ and $g(E_{\alpha\dot\alpha},E_{\beta\dot\beta})=\varepsilon_{\alpha\beta}\tilde\varepsilon_{\dot\alpha\dot\beta}$ with $\varepsilon_{\alpha\beta}=-\varepsilon_{\beta\alpha}$ and $\tilde\varepsilon_{\dot\alpha\dot\beta}=-\tilde\varepsilon_{\dot\beta\dot\alpha}$. In the presence of a contact metric structure, we have that the $\omega_\pm\in\Omega_\pm^2(M)$ are given in terms of the Hodge star $\star_5$ on $M$ by the formul{\ae}
\begin{equation}
 \omega_\pm\ =\ \tfrac12\big(\omega_H\pm\xi\lrcorner(\star_5\omega_H)\big)
 \end{equation} 
for a fixed orientation on $M$ and then, imposing $\omega\in\Omega^2_\pm(M)$ amounts to saying that
\begin{equation}
  \omega\ =\ \pm\star_5\!(\eta\wedge\omega)\ =\ \pm\xi\lrcorner(\star_5\omega)~.
\end{equation}

Finally, we call a conformal contact metric manifold $M$ a conformal $K$-contact manifold provided the Reeb vector field is Killing. 

\paragraph{Cauchy--Riemann twistor space.} 
After having presented the setup, we are now in the position to  discuss the twistor geometry of contact manifolds.  To this end, let $M$ be a five-dimensional conformal $K$-contact manifold with  Levi-Civita connection $\nabla$. Because of the factorisation $\sigma:H_\FC\overset{\cong}{\to} S\otimes\tilde S$, the connection $\nabla^H$ introduced in \eqref{eq:nablaH} is of the form $\nabla^H=\nabla^{H,S}\otimes\mbox{id}^{\tilde S}+\mbox{id}^S\otimes\nabla^{H,\tilde S}$, where `id' denotes the identity on the respective bundles.

Similarly to the four-dimensional setting, we wish to define the twistor space $Z$ of $M$ to be the projectivisation\footnote{This is essentially Itoh's definition \cite{Itoh:2002,Itoh:2003}. He first defined the twistor space as the space of anti-self-dual two-forms on the contact distribution of a certain length but later on argued that this definition is basically \eqref{eq:TS} (he considered the projectivisation of $\tilde S$ instead of $\tilde S^\vee$). We emphasise again that even when $\tilde S^\vee$ does not exist globally, its projectivisation does.}
\begin{equation}\label{eq:TS}
  Z\ :=\ \PP(\tilde S^\vee)
\end{equation}
of the dual of $\tilde S$.  This is a bundle over $M$ with projection $\pi:Z\to M$ with complex projective lines $\FC P^{1}$ as typical fibres. We shall endow these fibres with homogeneous fibre coordinates $\pi_{\dot\alpha}$, in the following.  

Clearly, since $Z$ is an odd-dimensional manifold, it cannot be equipped with an almost complex structure. However, it does come with a natural almost Cauchy--Riemann structure. Recall that an almost Cauchy--Riemann structure on a smooth manifold is a complex distribution in the complexified tangent bundle that does not contain any real vector fields. To introduce such a structure on $Z$, we first point out that with the help of $\nabla^H$, any vector field $X$ on $M$ is (horizontally) lifted to the twistor space $Z$ as
\begin{equation}\label{eq:lift}
 X'\ =\ X +X\lrcorner\omega^H{}_{\dot\alpha}{}^{\dot\beta}\left(\pi_{\dot\beta}\,\der{\pi_{\dot\alpha}}+\hat\pi_{\dot\beta}\,\der{\hat\pi_{\dot\alpha}}\right),
\end{equation}
where $\omega^H{}_{\dot\alpha}{}^{\dot\beta}$ denotes the connection one-form of $\nabla^{H,\tilde S}$ and
$\hat\pi_{\dot\alpha}$ is defined by the real structure $\tilde\tau$ (see e.g.~\cite{Woodhouse:1985id})
\begin{equation}
 \tilde\tau\ :\ (\pi_{\dot\alpha})\ =\ \begin{pmatrix} \pi_{\dot 1}\\ \pi_{\dot 2}\end{pmatrix}\ \mapsto\ 
(\hat \pi_{\dot\alpha})\ :=\ \begin{pmatrix}  -\pi^*_{\dot 2} \\ \pi^*_{\dot 1}\end{pmatrix}.
\end{equation}
Here, `$*$' denotes complex conjugation.  Using \eqref{eq:lift}, we may lift the frame fields $\{E_{\alpha\dot\alpha},\xi\}$ into $Z$. We shall also need the following complex vector fields
\begin{equation}\label{eq:LiftOfFrame}
 V_{\alpha\dot\alpha}\ :=\ E_{\alpha\dot\alpha} +E_{\alpha\dot\alpha}\lrcorner\omega^H{}_{\dot\beta}{}^{\dot\gamma}\,\pi_{\dot\gamma}\,\der{\pi_{\dot\beta}}\eand
  V_\xi\ :=\ \xi + \xi\lrcorner\omega^H{}_{\dot\alpha}{}^{\dot\beta}\,\pi_{\dot\beta}\,\der{\pi_{\dot\alpha}}
\end{equation}
that arise from the frame fields $\{E_{\alpha\dot\alpha},\xi\}$. These ingredients then allow us to introduce a natural distribution $D\hookrightarrow T_\FC Z$ in the complexified tangent bundle of $Z$ according to
\begin{equation}\label{eq:CRS}
 D\ :=\ \langle V_\alpha,V^{\dot\alpha}\rangle\ewith V_\alpha\ :=\ \pi^{\dot\alpha} V_{\alpha\dot\alpha}\eand V^{\dot\alpha}\ :=\ \der{\hat\pi_{\dot\alpha}}~,
\end{equation}
where $V_{\alpha\dot\alpha}$ is given in \eqref{eq:LiftOfFrame}. This distribution is of complex rank three since $\hat\pi_{\dot\alpha}V^{\dot\alpha}=0$. Furthermore, $D\cap D^*=0$. Hence, it defines an almost Cauchy--Riemann structure on $Z$. For that reason, we shall refer to $Z$ also as Cauchy--Riemann twistor space.\footnote{Here, we are following LeBrun's terminology \cite{LeBrun:1984} used for three-dimensional manifolds.} Now the question is under what circumstances the distribution \eqref{eq:CRS} can become integrable. The answer is given by the following theorem:\footnote{Itoh  \cite{Itoh:2002,Itoh:2003} proved this by translating the problem to conditions on the principal bundle of orthonormal frames on $H$ while here instead, we shall give a direct and simplified alternative proof by using \eqref{eq:CRS} and working with the bundle $\tilde S^\vee$ and its projectivisation---this is in spirit of Atiyah, Hitchin \& Singer's original treatment in the four-dimensional case \cite{Atiyah:1978wi} (see also Woodhouse \cite{Woodhouse:1985id}). Importantly, our approach will eventually enable us to write down a Lax pair for the contact instanton equation in the next section.}

\vspace{10pt}
{\theorem\label{thm:CRS}
Let $M$ be a five-dimensional conformal $K$-contact manifold with Levi-Civita connection $\nabla$ and curvature $\CR$. Consider the restriction of $\CR$ to $\Omega_-^2(M)$, that is, $\CR_-^-:\Omega_-^2(M)\to \Omega_-^2(M)$. The almost Cauchy--Riemann structure \eqref{eq:CRS} on the Cauchy--Riemann twistor space \eqref{eq:TS} of $M$ is integrable if and only if the totally trace-free part of $\CR_-^-$ vanishes.
}
 
\vspace{15pt}
\noindent
{\it Proof:} Using the frame fields $\{E_{\alpha\dot\alpha},\xi\}$, the totally trace-free part of $\CR_-^-$ is given by
\begin{equation}
 \CR_{\alpha(\dot\alpha\,\beta\dot\beta\,\gamma\dot\gamma\,\delta\dot\delta)}\ =\ 
 g\big(\CR(E_{\alpha(\dot\alpha},E_{\beta\dot\beta})E_{\gamma\dot\gamma},E_{\delta\dot\delta)}\big)~,
\end{equation}
where the parentheses indicate total normalised symmetrisation of all the enclosed dotted indices. Furthermore,  in the case of conformal $K$-contact manifolds, the Gau{\ss} equation \eqref{eq:Gauss} simplifies to 
\begin{equation}\label{eq:GaussK}
 g(\CR^H(X,Y)Z,W)\ =\ g(\CR(X,Y)Z,W)+\dd\eta(Z,X)\dd\eta(W,Y)-\dd\eta(Z,Y)\dd\eta(W,X)
\end{equation}
whenever $W\in \Gamma(M,H)$. Using the fact that $\dd\eta\in\Omega^2_+(M)$, we conclude that 
\begin{equation}
 \CR^H_{\alpha(\dot\alpha\,\beta\dot\beta\,\gamma\dot\gamma\,\delta\dot\delta)}\ =\ \CR_{\alpha(\dot\alpha\,\beta\dot\beta\,\gamma\dot\gamma\,\delta\dot\delta)}~.
\end{equation}
Since $\CR^H$ is of the form  $\CR^H=\CR^{H,S}\otimes\mbox{id}^{\tilde S}+\mbox{id}^S\otimes \CR^{H,\tilde S}$, the components $\CR^H_{\alpha(\dot\alpha\,\beta\dot\beta\,\gamma\dot\gamma\,\delta\dot\delta)}$ represent the totally trace-free part of $\CR^{H,\tilde S}_-$. Thus, we have $\CR^H_{\alpha(\dot\alpha\,\beta\dot\beta\,\gamma\dot\gamma\,\delta\dot\delta)}=\varepsilon_{\alpha\beta}\varepsilon_{\gamma\delta}\CC^H_{\dot\alpha\dot\beta\dot\gamma\dot\delta}$ with $\CC^H_{\dot\alpha\dot\beta\dot\gamma\dot\delta}$ being totally symmetric. Altogether, we arrive at 
\begin{equation}\label{eq:TFPR-C}
 \CR_{\alpha(\dot\alpha\,\beta\dot\beta\,\gamma\dot\gamma\,\delta\dot\delta)}\ =\ \varepsilon_{\alpha\beta}\varepsilon_{\gamma\delta}\CC^H_{\dot\alpha\dot\beta\dot\gamma\dot\delta}~.
\end{equation}

All that is is left now to show is that the integrability of $D$ is  equivalent to the vanishing of $\CC^H_{\dot\alpha\dot\beta\dot\gamma\dot\delta}$. This, however, is essentially the same statement as in the four-dimensional setting. To verify the integrability,  we simply have to compute the commutators of the vector fields generating the distribution. The only non-vanishing commutator is $[V_\alpha,V_\beta]$. To compute it, we use the formula
\begin{equation}
 [V_{\alpha\dot\alpha},V_{\beta\dot\beta}]-f_{\alpha\dot\alpha\,\beta\dot\beta}{}^{\gamma\dot\gamma}V_{\gamma\dot\gamma}-f_{\alpha\dot\alpha\,\beta\dot\beta}{}^{\xi}\,V_\xi\ =\ \CR^{H,\tilde S}_{\alpha\dot\alpha\,\beta\dot\beta\,\dot\gamma\,\dot\delta}\,\pi^{\dot\gamma}\der{\pi_{\dot\delta}}~,
\end{equation}
where the $f$s are the structure functions for the frame fields $E_{\alpha\dot\alpha}$ and $\xi$. Then
 \begin{equation}
 \begin{aligned}
  {[V_\alpha,V_\beta]}\ &=\ -\pi^{\dot\alpha}(E_{\alpha\dot\alpha}\lrcorner\omega^H{}_\beta{}^\gamma-E_{\beta\dot\alpha}\lrcorner\omega^H{}_\alpha{}^\gamma) V_\gamma\,+\\
  &\kern2cm+\,\pi^{\dot\alpha}\pi^{\dot\beta}\,\eta([E_{\alpha\dot\alpha},E_{\beta\dot\beta}])V_\xi-
     \varepsilon_{\alpha\beta}\pi^{\dot\alpha}\pi^{\dot\beta}\pi^{\dot\gamma}\CC^H_{\dot\alpha\dot\beta\dot\gamma\dot\delta}\der{\pi_{\dot\delta}}~,
\end{aligned}
 \end{equation}
where  $\omega^H{}_\alpha{}^\beta$ is the connection one-form of $\nabla^{H,S}$. To arrive at this expression, we used that the torsion  \eqref{eq:TorsionH} of $\nabla^H$ has components only along $\xi$. We also used the contact metricity. Next, with the help of 
\begin{equation}\label{eq:deta}
 \dd\eta(X,Y)\ =\ \tfrac12\big(X\eta(Y)-Y\eta(X)-\eta([X,Y])\big)~,
\end{equation}
we see that  $\eta([E_{\alpha(\dot\alpha},E_{\beta\dot\beta)}])=0$ since $\dd\eta\in\Omega^2_+(M)$. Therefore, the commutator $[V_\alpha,V_\beta]$ is given by 
\begin{equation}\label{eq:FinalCV}
{[V_\alpha,V_\beta]}\ =\ -\pi^{\dot\alpha}(E_{\alpha\dot\alpha}\lrcorner\omega^H{}_\beta{}^\gamma-E_{\beta\dot\alpha}\lrcorner\omega^H{}_\alpha{}^\gamma) V_\gamma-\varepsilon_{\alpha\beta}\pi^{\dot\alpha}\pi^{\dot\beta}\pi^{\dot\gamma}\CC^H_{\dot\alpha\dot\beta\dot\gamma\dot\delta}\der{\pi_{\dot\delta}}~,
\end{equation}
and we may conclude that $D$ is integrable if and only if $\CC^H_{\dot\alpha\dot\beta\dot\gamma\dot\delta}=0$. By virtue of \eqref{eq:TFPR-C}, this is equivalent to saying that the totally  trace-free part of $\CR_-^-$ vanishes. This completes the proof. \hfill$\Box$

\vspace{15pt}
{\remark\label{rem:ext}
It is clear from the proof that the only place where we used the Killing property of the  Reeb vector field is in the Gau{\ss} equation \eqref{eq:GaussK}. Since the integrability condition $\CC^H_{\dot\alpha\dot\beta\dot\gamma\dot\delta}=0$ does not depend on property of  $\xi$ being Killing, we may relax that condition and simply consider conformal contact metric  manifolds as a generalisation of the above theorem. In that case, we need to work with the general Gau{\ss} equation \eqref{eq:Gauss}, and the requirement
\begin{equation}\label{eq:CurvCMM}
  \CR_{\alpha(\dot\alpha\,\beta\dot\beta\,\gamma\dot\gamma\,\delta\dot\delta)}\ =\ \big[g(E_{\gamma\dot\gamma},\nabla_{\alpha\dot\alpha}\xi)\,g(E_{\delta\dot\delta},\nabla_{\beta\dot\beta}\xi)-
  g(E_{\gamma\dot\gamma},\nabla_{\beta\dot\beta}\xi)\,g(E_{\delta\dot\delta},\nabla_{\alpha\dot\alpha}\xi)\big]_{(\dot\alpha\dot\beta\dot\gamma\dot\delta)}~.
\end{equation}
replaces that of the vanishing of $\CR_{\alpha(\dot\alpha\,\beta\dot\beta\,\gamma\dot\gamma\,\delta\dot\delta)}$ in the $K$-contact case.
}

 \paragraph{\mathversion{bold}$F$-structures.}
 Next we wish to introduce a particular $F$-structure on the twistor space which eventually allows us to discuss the contact instanton equations via a Penrose--Ward transform.\footnote{Such structures appear in various twistor constructions. For instance, the twistorial description of the three-dimensional Bogomolny monopole equation (and its supersymmetric extension) in terms of real geometries involves $F$-structures quite naturally \cite{Popov:2005uv}.} Recall that  Rawnsley \cite{Rawnsley} defined an almost $F$-structure on a smooth manifold $M$ to be a distribution $F$ in the complexified tangent bundle of $M$ such that $F\cap F^*$ has constant rank. Hence, an almost Cauchy--Riemann structure is a special instance of an almost $F$-structure when $F\cap F^*=0$. An almost $F$-structure is called an $F$-structure if in addition both $F$ and $F\cap F^*$ are integrable. In that case, the almost $F$-structure is said to be integrable.
 
On contact manifolds we have a natural vector field, the Reeb vector field. Moreover, we have seen above how it lifts into the twistor space. We have therefore a natural candidate of an almost $F$-structure on $Z$ defined by
\begin{equation}\label{eq:FS}
 F\ :=\ \langle V_\alpha,\,\xi',\,  V^{\dot\alpha}\rangle~,
\end{equation}
where $V_{\alpha}$ and $V^{\dot\alpha}$ were given in \eqref{eq:CRS} and $\xi'$ is the lift of the Reeb vector field via \eqref{eq:lift}. Notice that upon action on functions holomorphic in the $\pi_{\dot\alpha}$-coordinates, the vector field $\xi'$ reduces to $V_\xi$ given in \eqref{eq:LiftOfFrame}. Clearly,  $F\cap F^*$ is of constant rank one and hence is integrable. Therefore, the integrability of the almost $F$-structure boils down to the integrability of the distribution $F$. The following theorem tells us when this is happening:
 
\vspace{10pt}
{\theorem\label{thm:FS}
Let $M$ be a five-dimensional conformal $K$-contact manifold with Levi-Civita connection $\nabla$ and curvature $\CR$. Consider the restrictions of $\CR$ to $\Omega_-^2(M)$ and $\eta\wedge\Omega^1_H(M)$, respectively, that map into $\Omega_-^2(M)$, i.e.~$\CR_-^-:\Omega_-^2(M)\to \Omega_-^2(M)$ and $\CR_\eta^-:\eta\wedge\Omega^1_H(M)\to \Omega_-^2(M)$. The almost $F$-structure \eqref{eq:FS} on the Cauchy--Riemann twistor space \eqref{eq:TS} of $M$ is integrable if and only if the totally trace-free parts of $\CR_-^-$ and $\CR_\eta^-$ vanish.\footnote{Notice that if one only requires the totally trace-free part of $\CR_\eta^-$ to vanish then this is the condition needed for $\xi'$ to be a Cauchy--Riemann vector field \cite{Itoh:2002,Itoh:2003}.}
} 

\vspace{15pt} 
\noindent
{\it Proof:} We need to compute the commutators of the vector fields generating the distribution $F$. The only non-vanishing ones are $[V_\alpha,V_\beta]$, $[V^{\dot\alpha},\xi']$ and $[V_\alpha,\xi']$.  The first one was already computed when proving Theorem \ref{thm:CRS} and is given in \eqref{eq:FinalCV}. The second one is proportional to $V^{\dot\alpha}$ and hence does not give any conditions. To compute the last commutator, we first note that
\begin{equation}
 [V_{\alpha\dot\alpha},V_\xi] -f_{\alpha\dot\alpha\,\xi}{}^{\beta\dot\beta}V_{\beta\dot\beta}-f_{\alpha\dot\alpha\,\xi}{}^{\xi}\,V_\xi\ =\ \CR^{H,\tilde S}_{\alpha\dot\alpha\,\xi\,\dot\gamma\,\dot\delta}\,\pi^{\dot\gamma}\der{\pi_{\dot\delta}}~.
\end{equation}
This simplifies since $f_{\alpha\dot\alpha\,\xi}{}^{\xi}=0$ which follows directly from \eqref{eq:deta}.  Moreover, $\dd\eta\in\Omega^2_+(M)$ implies that the endomorphism $\phi|_H$ has $\phi_{\alpha\dot\alpha}{}^{\beta\dot\beta}=\phi_\alpha{}^\beta\delta_{\dot\alpha}{}^{\dot\beta}$ as components in the $E_{\alpha\dot\alpha}$-basis. Using $\nabla_X\xi=-\phi(X)$ and the fact that the torsion  \eqref{eq:TorsionH} of $\nabla^H$ has components only along $\xi$, we arrive after some straightforward algebraic manipulations at
\begin{equation}\label{eq:FinalCVxi}
 [V_\alpha,\xi']\ =\ -(\phi_\alpha{}^\beta+\xi\lrcorner\omega^H{}_\alpha{}^\beta)V_\beta-\pi^{\dot\alpha}\pi^{\dot\beta}\CR^{H,\tilde S}_{\alpha(\dot\alpha\,\xi\,\dot\beta\dot\gamma)}\der{\pi_{\dot\gamma}}
 +\pi^{\dot\alpha}E_{\alpha\dot\alpha}(\xi\lrcorner\omega^H{}_{\dot\beta}{}^{\dot\gamma})\hat\pi_{\dot\gamma}\der{\hat\pi_{\dot\beta}}~.
\end{equation}
Furthermore, the Gau{\ss} equation \eqref{eq:GaussK} directly implies that $\varepsilon_{\beta\gamma}\CR^{H,\tilde S}_{\alpha\dot\alpha\,\xi\,\dot\beta\dot\gamma}=\CR_{\alpha\dot\alpha\,\xi\,\beta\dot\beta\,\gamma\dot\gamma}$. By symmetrising the dotted indices one obtains the totally trace-free part of $\CR_\eta^-$. Altogether,
\begin{equation}
 \CR_{\alpha(\dot\alpha\,\beta\dot\beta\,\gamma\dot\gamma\,\delta\dot\delta)}\ =\ \varepsilon_{\alpha\beta}\varepsilon_{\gamma\delta}\CC^{H}_{\dot\alpha\dot\beta\dot\gamma\dot\delta}\eand 
 \CR_{\alpha(\dot\alpha\,\xi\,\beta\dot\beta\,\gamma\dot\gamma)}\ =\ \varepsilon_{\beta\gamma}\CR^{H,\tilde S}_{\alpha(\dot\alpha\,\xi\,\dot\beta\dot\gamma)}
\end{equation} 
and from \eqref{eq:FinalCV} and \eqref{eq:FinalCVxi} we may conclude that the $F$-structure is integrable if and only if these curvature components vanish. \hfill$\Box$

\vspace{10pt}
{\remark
As noted previously, Sasakian manifolds can be defined by the curvature equation \eqref{eq:CurvSas}. This condition immediately implies that $\CR_{\alpha\dot\alpha\,\xi\,\beta\dot\beta\,\gamma\dot\gamma}=0$ such that  the totally trace-free part of $\CR_\eta^-$ clearly vanishes. Thus, Sasakian manifolds that obey the curvature condition of Theorem \ref{thm:CRS} provide examples on which the almost $F$-structure \eqref{eq:FS} is integrable.
}
 
{\remark\label{rem:extF}
In Remark \ref{rem:ext}, we have explained that  the condition for the almost Cauchy--Riemann structure $D$ to be integrable does not really depend on the property of $\xi$ being Killing, thus allowing us to generalise Theorem \ref{thm:CRS} to arbitrary contact metric manifolds with \eqref{eq:CurvCMM}. In contrast, the proof of Theorem \ref{thm:FS} makes use of the equation  $\nabla_X\xi=-\phi(X)$ which is the Killing condition on $\xi$. At the moment, it is not clear if one can generalise this to arbitrary contact metric (modulo curvature conditions) in a sensible way. 
} 
 
\paragraph{Higher-dimensional extension.}
Bailey \& Eastwood   \cite{Bailey:1991xx} gave the notion of a paraconformal structure as a higher-dimensional generalisation of a conformal structure on four-dimensional spin manifolds. In particular, if $M$ is a smooth manifold of dimension $pq$, then a $(p,q)$-paraconformal structure on $M$ is a factorisation of the complexified tangent bundle $\sigma:T_\FC M\overset{\cong}{\to} S\otimes\tilde S$ into two complex vector bundles $S$ and $\tilde S$ of respective ranks  $p$ and $q$, and a fixed isomorphism $\det S\cong\det \tilde S$ of the corresponding determinant line bundles. In addition, one also assumes that the bundles $S$ and $\tilde S$ come equipped with real structures $\tau:S\to S$ and $\tilde\tau:\tilde S\to\tilde S$ which are of the previous type, i.e.~they are anti-linear and obey the conditions $\tau^2=-1=\tilde\tau^2$ such that i) the real structure $\tau_M$ induced on $T_\FC M$ is given by $\tau_M=\tau\otimes\tilde\tau$ and obeys $\tau_M^2=1$ and ii) the set of fixed points of $\tau_M$ coincides with $TM$. 
 
In the present context of a $(2m+1)$-dimensional contact manifold $M$, we wish to change this point of view slightly and introduce a paraconformal structure only on the contact distribution $H$. Since the structure group for contact manifolds can be reduced to $\sU(m)\times 1$, a suitable assumption is  a factorisation of the form $\sigma:H_\FC\overset{\cong}{\to} S\otimes\tilde S$, where rk$\,S=2$ and rk$\,\tilde S=m\in 2\NN$ together with the indentification  $\det S\cong\det \tilde S$ and the real structures $\tau$ and $\tilde\tau$ of the above type. We shall refer to this type of structure as a $(2,m)$-paraconformal contact structure and to contact manifolds equipped with such a structure as $(2,m)$-paraconformal contact manifolds.\footnote{One might also call them quaternionic contact manifolds, but since we shall consider the connection $\nabla^H$ which is not torsion-free, we feel that the above terminology (despite being longer) is more suitable. }  Note that we then have the same decompositions \eqref{eq:Decompo} and \eqref{eq:TwoForms} 

Next we call $(2,m)$-paraconformal contact manifold $M$  contact metric if the datum $(\eta,\xi,g,\phi)$ satisfies the relations \eqref{eq:CMS} and is compatible with the paraconformal structure, that is, the endomorphism $\phi$ and the metric $g$ restrict on $H$ to $\phi|_H=\phi^S\otimes\mbox{id}^{\tilde S}$  and $g|_H=\varepsilon\otimes\tilde\varepsilon$, where $\varepsilon\in\Gamma_\tau(M,\Lambda^2 S^\vee)$ and $\tilde\varepsilon\in \Gamma_{\tilde\tau}(M,\Lambda^2\tilde S^\vee)$ are both of maximal rank. Thus, $\dd\eta\in\Omega^2_+(M)$.  In addition, we call a  $(2,m)$-paraconformal contact metric manifold $M$ a  $(2,m)$-paraconformal $K$-contact manifold provided the Reeb vector field is a Killing vector field. 
 
We may now proceed as in previous paragraphs, and define the Cauchy--Riemann twistor space to be the projectivisation of the dual of $\tilde S$, that is, $Z:=\PP(\tilde S^\vee)$. Now this is a $\CPP^{m-1}$-bundle over $M$ which is of real dimension $4m-1$. Then there are again natural almost Cauchy--Riemann   and $F$-structures  whose form is basically the same as in \eqref{eq:CRS} and \eqref{eq:FS}, i.e.
\begin{subequations}
\begin{equation}
 D\ :=\ \langle V^\alpha, V^{\dot\alpha}\rangle
 \eand
 F\ :=\ \langle V^\alpha,\xi',V^{\dot\alpha}\rangle
 \end{equation} 
with
\begin{equation}
\begin{aligned}
 V^{\alpha}\ :=\ \pi_{\dot\alpha}\left(E^{\alpha\dot\alpha} +E^{\alpha\dot\alpha}\lrcorner\omega^H{}_{\dot\beta}{}^{\dot\gamma}\,\pi_{\dot\gamma}\,\der{\pi_{\dot\beta}}\right),\qquad
 V^{\dot\alpha}\ :=\ \der{\hat\pi_{\dot\alpha}}~,\\
 \\[-10pt]
  \xi'\ :=\ \xi +\xi\lrcorner\omega^H{}_{\dot\alpha}{}^{\dot\beta}\left(\pi_{\dot\beta}\,\der{\pi_{\dot\alpha}}+\hat\pi_{\dot\beta}\,\der{\hat\pi_{\dot\alpha}}\right),\kern1.2cm
  \end{aligned}
 \end{equation}
 \end{subequations}
where $\alpha,\beta,\ldots=1,2$ but $\dot\alpha,\dot\beta,\ldots=1,\ldots,m$ and $E^{\alpha\dot\alpha}=g^{\alpha\dot\alpha\beta\dot\beta}E_{\beta\dot\beta}$ and $\hat\pi_{\dot\alpha}=\tilde\tau(\pi_{\dot\alpha})$. Notice that the distribution $D$ that determines the almost Cauchy--Riemann structure is of complex rank $m+1\leq2m-1$ which is less than the maximal possible rank---the almost Cauchy--Riemann structure is of hypersurface type only when $m=2$. The integrability of these structures is again determined by the vanishing of the totally trace-free parts of the curvature components $\CR_-^-$ and $\CR_\eta^-$ introduced in Theorem \ref{thm:CRS} and Theorem \ref{thm:FS}, so we may simply replace the phrase ``five-dimensional conformal $K$-contact manifold" by the phrase ``$(2,m)$-paraconformal $K$-contact manifold" in these theorems. The proofs go through without alteration.\footnote{Helpful in verifying the assertions is also the appendix of Bailey \& Eastwood's paper \cite{Bailey:1991xx}, where details about the decomposition of the curvature (in our context $\CR^H$) into irreducible pieces are given.}

\vspace{5pt}
{\theorem\label{thm:FS-para}
Let $M$ be a $(2,m)$-paraconformal $K$-contact manifold with Levi-Civita connection $\nabla$ and curvature $\CR$. Let $Z$ be its twistor space equipped with the almost  Cauchy--Riemann structure $D=\langle V^\alpha, V^{\dot\alpha}\rangle$ and the almost $F$-structure $F=\langle V^\alpha,\xi',V^{\dot\alpha}\rangle$. Then $D$ is integrable if and only if the totally trace-less part of $\CR_-^-:\Omega_-^2(M)\to \Omega_-^2(M)$ vanishes while the $F$-structure is integrable if, in addition, the totally trace-less part of $\CR_\eta^-:\eta\wedge\Omega^1_H(M)\to \Omega_-^2(M)$ vanishes, as well.  
} 

\vspace{15pt}
Particularly interesting is the case when the paraconformal structure is of the form $\sigma:H_\FC\overset{\cong}{\to} S\otimes \odot^q\tilde S$ for two rank-$2$ bundles $S$ and $\tilde S$ and $q+1=m$.\footnote{For twistor constructions that use paraconformal structures on smooth manifolds $M$ of the form $T_\FC M\cong\odot^q S$ for some rank-$2$ vector bundle $S$, see e.g.~\cite{Ward:1983zm,LeBrun:1984,Dunajski:2005}.} This reduces the structure group $\sU(m)\times 1$ further down to  $\sU(2)\times 1$ which is the same as in the five-dimensional case. We shall come back to this case in the next section, where we explain that the Penrose--Ward transform for such paraconformal $K$-contact manifolds gives rise to certain higher-dimensional contact instanton equations in spirit of Ward's self-dual models \cite{Ward:1983zm}. Finally, we would like to mention that Vezzoni \cite{Vezzoni:2010} introduced a twistor space for higher-dimensional contact manifolds which, however, appears to differ from ours presented above.

\section{Penrose--Ward transform and contact instantons}\label{sec:PWtrafo}

As we shall explain in this section, the Ward construction \cite{Ward:1977ta,Atiyah:1978wi}\footnote{Itoh \cite{Itoh:2002,Itoh:2003} already mentioned in passing the possibility of an extension of the Ward construction.} of four-dimensional Yang--Mills instantons can be naturally extended to the present case of contact instantons, 
\begin{equation}\label{eq:CIE-2}
  \CF\ =\ \star_5(\eta\wedge\CF)\ =\ \xi\lrcorner(\star_5\CF)~,
\end{equation}
that is, we shall establish certain types of Penrose--Ward transforms. The basic idea is to use so-called partial or $F$-connections \cite{Rawnsley}. 

\paragraph{\mathversion{bold}$F$-connection.}
Let us start by recalling the notion of an $F$-connection in the sense of Rawnsley \cite{Rawnsley}. Let $M$ be a smooth manifold with an $F$-structure. For any smooth function $f$ on $M$, let $\dd_F f$ be the restriction of the exterior derivative $\dd f$ to $F$, i.e.~$\dd_F$ is the composition $C^\infty(M)\overset{\dd}{\to} \Gamma(M,T_\FC^\vee M)\to\Gamma(M,F^\vee)$. We shall write $\Omega^k_F(M):=\Gamma(M,\Lambda^k F^\vee)$ in the following. Elements of $\Omega^k_F(M)$ are called relative differential $k$-forms. Note that we can extend $\dd_F$ to act on relative $k$-forms, $\dd_F:\Omega^k_F(M)\to \Omega^{k+1}_F(M)$. Let now $E$ be a complex vector bundle over $M$. A connection along the distribution $F$ is called an $F$-connection, $\partial_{\CA_F}:E\to\Omega^1_F(M,E)$, provided it satisfies the Leibniz rule $\partial_{\CA_F}(fs)=(\dd_F f)s+f\partial_{\CA_F}s$, where $s$ is a section of $E$, $f$ a function on $M$, and $\Omega^k_F(M,E):=\Omega^k_F(M)\otimes E$. This extends to a connection $\partial_{\CA_F}:\Omega^k_F(M,E)\to\Omega^{k+1}_F(M,E)$. Locally, we have $\partial_{\CA_F}=\dd_F+\CA_F$, where $\CA_F$ is an End$\,E$-valued connection one-form  that has components only along $F$. As always, also the connection $\partial_{\CA_F}$ induces a curvature two-form $\CF_F:=\partial_{\CA_F}^2\in\Omega^{2}_F(M,\mbox{End}\,E)$, and the bundle $(E,\partial_{\CA_F})$ is called $F$-flat (or partially flat) provided $\CF_F=0$ on  $M$. Note that when the $F$-structure is a Cauchy--Riemann structure and  $\CF_F=0$,  then $(E,\partial_{\CA_F})$ is also called a Cauchy--Riemann vector bundle.

\paragraph{Penrose--Ward transform.}
Having recalled the definition of an $F$-connection, let us now construct Penrose--Ward transforms for vector bundles over contact manifolds. For the moment, let us focus on five-dimensional contact manifolds. Then we have:\footnote{Notice that one could relax the Killing property and work with general contact metric manifolds with \eqref{eq:CurvCMM} since the integrability of the almost Cauchy--Riemann structure \eqref{eq:CRS} does not depend on it; see Remark \ref{rem:ext}.}

\vspace{10pt}
{\theorem\label{thm:CRS-PWtrafo}
Let $M$ be a five-dimensional $K$-contact manifold with Cauchy--Riemann twistor space $\pi:Z\to M$ as in  \eqref{eq:TS} and (integrable) Cauchy--Riemann structure \eqref{eq:CRS}. There is a one-to-one correspondence between
\begin{itemize}\setlength{\itemsep}{-1mm}
\item[{(i)}] rank-$r$ Cauchy--Riemann vector bundles $E_Z\to Z$ such that the restriction $E_Z|_{\pi^{-1}(p)}$ is holomorphically trivial for all $p\in M$ and
\item[{(ii)}] rank-$r$ complex vector bundles $E_M\to M$ equipped with a connection $\partial_{\CA}$ and curvature $\CF=\partial_{\CA}^2$ such that the projection on the contact distribution is $\CF_H\in\Omega^2_+(M,\operatorname{End}E_M)$, that is, $\CF_-=0$.
\end{itemize}  
}

\vspace{15pt}
\noindent
{\it Proof:}
Let $E_Z$ be a rank-$r$ complex vector bundle over $Z$ that is $D$-flat with respect to the distribution   \eqref{eq:CRS} and holomorphically trivial up the fibres of $Z\to M$. Then there exist $r$ linearly independent  sections $s_a$ of $E_Z$ and $a=1,\ldots,r$, which are covariantly constant with respect to $\partial_{\CA_D}=\dd_D+\CA_D$. In addition, there exists a gauge of $\CA_D$ in which the sections $s_a$ become holomorphic in the fibre coordinates $\pi_{\dot\alpha}$. Explicitly, we then have
\begin{equation}\label{eq:SectionsE}
  (V_\alpha+\CA_\alpha)s_a\ =\ 0\eand V^{\dot\alpha}s_a\ =\ 0\ewith \CA_\alpha\ :=\ V_\alpha\lrcorner\CA_D~.
\end{equation}
Then it is rather easy to see that $V^{\dot\alpha}\CA_\alpha=0$ and hence, $\CA_{\alpha}$ must be of homogeneity degree one in $\pi_{\dot\alpha}$. Therefore, we may write $\CA_{\alpha}=:\pi^{\dot\alpha}\CA_{\alpha\dot\alpha}$, where $\CA_{\alpha\dot\alpha}$ does not depend on the fibre coordinates and is defined locally on $M$. Then  $\CA_{\alpha\dot\alpha}$ is interpreted as component $\CA_{\alpha\dot\alpha}:=E_{\alpha\dot\alpha}\lrcorner\CA$ of a connection one-form $\CA$ on $M$. Notice that the component  $\CA_\xi:=\xi\lrcorner\CA$ is not fixed in this construction. Notice also that \eqref{eq:SectionsE} is invariant under the residual gauge transformations $s_a\to g^{-1}s_a$ and $\CA_\alpha\mapsto\ g^{-1} (V_\alpha+\CA_\alpha)g$ for $\sGL(r,\FC)$-valued $g$ that obey $V_{\dot\alpha}g=0$ and thus, $g$ must be constant up the fibres. Such $g$ mediate precisely the gauge transformations of  $\CA$ on $M$. In summary, we find a vector bundle $E_M\to M$ whose fibre at $p\in M$ is the space of holomorphic sections of $E_Z|_{\pi^{-1}(p)}$ which is equipped with a connection $\partial_\CA$.

To compute $\CF=\partial_\CA^2$, we need to find the compatibility conditions of \eqref{eq:SectionsE}. Using, \eqref{eq:FinalCV} and the fact that the torsion  \eqref{eq:TorsionH} has components only along the Reeb vector field $\xi$, we obtain 
\begin{equation}
 \pi^{\dot\alpha}\pi^{\dot\beta}\big(E_{\alpha\dot\alpha}\CA_{\beta\dot\beta}-E_{\beta\dot\beta}\CA_{\alpha\dot\alpha}+[\CA_{\alpha\dot\alpha},\CA_{\beta\dot\beta}]-f_{\alpha\dot\alpha\,\beta\dot\beta}{}^{\gamma\dot\gamma}\CA_{\gamma\dot\gamma}\big)\ =\ 0
\end{equation}
after some algebraic manipulations. Since the expression in the parentheses does not depend on $\pi_{\dot\alpha}$, it must vanish, and since $\eta([E_{\alpha(\dot\alpha},E_{\beta\dot\beta)}])=f_{\alpha(\dot\alpha\,\beta\dot\beta)}{}^{\xi}=0$ because $\dd\eta\in \Omega^2_+(M)$, we conclude that $\CF_-=0$. Altogether, we obtain a rank-$r$ complex vector bundle $E_M\to M$ over $M$ with a connection $\partial_\CA$ with  connection one-form $\CA$ with  $\CA_{\alpha\dot\alpha}=E_{\alpha\dot\alpha} \lrcorner \CA$ as above while $\CA_\xi=\xi\lrcorner\CA$ is undetermined such that $\CF_H=(\partial_\CA^2)_H\in\Omega^2_+(M,\operatorname{End}E_M)$.

Conversely, given a rank-$r$ complex vector bundle $E_M\to M$ over $M$ with a connection $\partial_\CA$  such that $\CF_H\in\Omega^2_+(M,\operatorname{End}E_M)$, then we may define a rank-$r$ complex vector bundle $E_Z\to Z$ over twistor space whose trivialisation is given by  the solutions to \eqref{eq:SectionsE} with $\CA_{\alpha}=\pi^{\dot\alpha}\CA_{\alpha\dot\alpha}$. This bundle is $D$-flat and holomorphically trivial up the fibres. This concludes the proof. \hfill $\Box$

\vspace{15pt}
In order to have that $\CF\in\Omega^2_+(M,\operatorname{End}E_M)$ and not just $\CF_H\in\Omega^2_+(M,\operatorname{End}E_M)$, we need by virtue of \eqref{eq:TwoForms} that both $\CF_-$ and $\CF_\eta$ vanish.  Thus, to give a twistor construction of the contact instanton equation, we need to extend the above construction to enforce $\CF_\eta=0$, as well. This is achieved by working with the $F$-structure given in  \eqref{eq:FS}.

\vspace{10pt}
{\theorem\label{thm:FS-PWtrafo}
Let $M$ be a five-dimensional $K$-contact manifold with Cauchy--Riemann twistor space $\pi:Z\to M$ as in  \eqref{eq:TS} and (integrable) $F$-structure \eqref{eq:FS}.\footnote{Because of Remark \ref{rem:extF}, it is not clear at the moment if one can generalise this theorem by relaxing the Killing property.} Then there is a one-to-one correspondence between
\begin{itemize}\setlength{\itemsep}{-1mm}
\item[{(i)}] $F$-flat rank-$r$ complex vector bundles $E_Z\to Z$ such that the restriction $E_Z|_{\pi^{-1}(p)}$ is holomorphically trivial for all $p\in M$ and
\item[{(ii)}] rank-$r$ complex vector bundles $E_M\to M$ equipped with a connection $\partial_{\CA}$ and curvature $\CF=\partial_{\CA}^2$ such that $\CF\in\Omega^2_+(M,\operatorname{End}E_M)$, that is, $\CF_-=0=\CF_\eta$.
\end{itemize}  
}

\vspace{15pt}
\noindent
{\it Proof:}
The proof is very similar to the previous one. In the present case, the equations \eqref{eq:SectionsE} get extended to
\begin{equation}\label{eq:SectionsEF}
  (V_\alpha+\CA_\alpha)s_a\ =\ 0~,\quad (V_\xi+\CA_\xi)s_a\ =\ 0~,\eand V^{\dot\alpha}s_a\ =\ 0
  \end{equation}
with $\CA_\alpha:=V_\alpha\lrcorner\CA_F$ and $\CA_\xi:=\xi'\lrcorner\CA_F$. Here, we made use of the fact that when the $s_a$ are holomorphic in $\pi_{\dot\alpha}$, i.e.~when $ V^{\dot\alpha}s_a=0$, then $\xi's_a=V_\xi s_a$, where $V_\xi$ was given in  \eqref{eq:LiftOfFrame}. As before, $V^{\dot\alpha}\CA_\alpha=0$ but now we also have $V^{\dot\alpha}\CA_\xi=0$, so $\CA_\alpha$ is of homogeneity one while $\CA_\xi$ is of homogeneity zero (i.e.~it does not depend on $\pi_{\dot\alpha}$). From \eqref{eq:SectionsEF}, we obtain
\begin{equation}
\begin{aligned}
 \pi^{\dot\alpha}\pi^{\dot\beta}\big(E_{\alpha\dot\alpha}\CA_{\beta\dot\beta}-E_{\beta\dot\beta}\CA_{\alpha\dot\alpha}+[\CA_{\alpha\dot\alpha},\CA_{\beta\dot\beta}]-f_{\alpha\dot\alpha\,\beta\dot\beta}{}^{\gamma\dot\gamma}\CA_{\gamma\dot\gamma}\big)\ &=\ 0~,\\
 \pi^{\dot\alpha}\big(E_{\alpha\dot\alpha}\CA_{\xi}-\xi\CA_{\alpha\dot\alpha}+[\CA_{\alpha\dot\alpha},\CA_{\xi}]-f_{\alpha\dot\alpha\,\xi}{}^{\gamma\dot\gamma}\CA_{\gamma\dot\gamma}\big)\ &=\ 0~.
 \end{aligned}
\end{equation}
Since $f_{\alpha(\dot\alpha\,\beta\dot\beta)}{}^{\xi}=0=f_{\alpha\dot\alpha\,\xi}{}^{\xi}$, we may conclude that the
 $\CF_-$ and $\CF_\eta$ components of $\CF=\partial_{\CA}^2$ vanish. Therefore, we find a rank-$r$ complex vector bundles $E_M\to M$ equipped with a connection $\partial_{\CA}$ and curvature $\CF=\partial_{\CA}^2$ such that $\CF\in\Omega^2_+(M,\operatorname{End}E_M)$. \hfill $\Box$

\vspace{15pt}
{\remark Let $\psi$ be some matrix-valued function depending meromorphically on $\lambda\in\FC P^1$ (the spectral parameter). Then we may write down a more familiar form of an auxiliary linear system (Lax pair formulation)
\begin{subequations}
\begin{equation}
  \lambda^{\dot\alpha}(V_{\alpha\dot\alpha}+\CA_{\alpha\dot\alpha})\psi\ =\ 0\eand (V_\xi+\CA_\xi)\psi\ =\ 0~,
\end{equation}
where $(\lambda_{\dot\alpha}):=(1,\lambda)$ and
\begin{equation}
 V_{\alpha\dot\alpha}\ =\ E_{\alpha\dot\alpha} -E_{\alpha\dot\alpha}\lrcorner\omega^H{}_{\dot\beta}{}^{\dot\gamma}\,\lambda_{\dot\gamma}\lambda^{\dot\beta}\,\der{\lambda}\eand
  V_\xi\ =\ \xi - \xi\lrcorner\omega^H{}_{\dot\alpha}{}^{\dot\beta}\,\lambda_{\dot\beta}\lambda^{\dot\alpha}\,\der{\lambda}~.
\end{equation}
\end{subequations}
 The compatibility condition is the contact instanton equation.
}

{\remark Notice that we may introduce real structures on $E_M$ and $E_Z$ induced by $\tau$ and $\tilde \tau$
such that the connection one-form $\CA$ on $M$ takes values in $\mathfrak{u}(r)$. If one also requires that $\det E_Z$ and $\det E_M$ are trivial, then one can reduce $\mathfrak{u}(r)$ further to $\mathfrak{su}(r)$. }

{\remark
Having provided a Penrose--Ward transform and a Lax pair, we may use them to discuss the hidden symmetry structures of the contact instanton equation. In particular, one is interested in infinitesimal deformations of the vector bundle $E_Z\to M$ which preserve the $F$-structure. Using the techniques developed in \cite{Popov:1998pc,Wolf:2004hp,Wolf:2005sd,Popov:2006qu} (see also \cite{Pohlmeyer:1979ya,Dolan:1983bp,Popov:1995qb,Ivanova:9702144,Ward:1990vs,Mason:1991rf,Dunajski:2009aa}), one may infinitesimally deform the transition functions of $E_Z$ to obtain the linearisation of \eqref{eq:SectionsEF}. From these linear equations, one can then extract the deformations $\delta\CA_F$ of $\CA_F$ which in turn lead to deformations $\delta\CA$ of $\CA$ which obey the linearised contact instanton equation (in the background of $\CA$). For instance, this way one can quickly obtain Kac--Moody type symmetries associated with the structure group of $E_M$. 
}

{\remark
The contact form is not closed and hence,  \eqref{eq:CIE} does not automatically imply the Yang--Mills equation but rather the Yang--Mills equation with torsion $\partial_\CA\star_5\!\CF=\star_5\CH\wedge\CF$, where $\CH:=\star_5\dd\eta$ is the torsion three-form.\footnote{The torsion Yang--Mills equation appears naturally in string theory, see e.g.~\cite{Grana:2005jc,Douglas:2006es,Blumenhagen:2006ci} and references therein.} The term $\star_5\CH\wedge\CF=\dd\eta\wedge\CF$ is a four-form on the contact distribution since $\xi\lrcorner(\dd\eta\wedge\CF)=0$. It does not vanish automatically in our case, since both $\dd\eta$ and $\CF$ are in $\Omega^2_+(M)$. Notice that the present twistor setup leads by construction to the situation when  $\dd\eta\in\Omega^2_\pm(M)$ and  $\CF\in\Omega^2_\pm(M,\operatorname{End}E_M)$; for $\dd\eta\in\Omega_-^2(M)$ one can construct a similar  twistor space as $\PP(S^\vee)$. One may consider the situation when $\dd\eta\in\Omega^2_\pm(M)$  while $\CF\in\Omega^2_\mp(M,\operatorname{End}E_M)$ in which case $\dd\eta\wedge\CF=0$. This is the situation that was considered by  Harland \& N\"olle in the Sasaki--Einstein setting and for $\sSU(2)$ as gauge group \cite{Harland:2011zs}. Therefore, we may conclude that the contact instanton equation with $\dd\eta\in\Omega^2_\pm(M)$ and  $\CF\in\Omega^2_\pm(M,\operatorname{End}E_M)$ appears to be integrable via the above construction but does not automatically imply the (torsion-free) Yang--Mills equation while the contact instanton equation with $\dd\eta\in\Omega^2_\pm(M)$ and  $\CF\in\Omega^2_\mp(M,\operatorname{End}E_M)$ implies the (torsion-free) Yang--Mills equation but it remains to be seen if this case is integrable in general.\footnote{If we consider a maximally degenerate contact structure so that $\dd\eta=0$, we have $M\to S^1$ or $M\to\FR$. In this case, both type of contact instanton equations are integrable and both imply the Yang--Mills equation.} 

}

{\remark\label{rmk:action}
 In general, one might consider instanton equations on a $d$-dimensional manifold of the form $\CF=\star_d(\Sigma\wedge\CF)$, where $\Sigma$ is a $(d-4)$-form. The resulting Yang--Mills equation with torsion, $\partial_\CA{\star_d\CF}=\dd\Sigma\wedge\CF$,  can be obtained from an action functional  \cite{Harland:2010ix,Bauer:2010fia}. Via a Bogomolny argument, solutions to $\CF=\star_d(\Sigma\wedge\CF)$ are in turn the absolute minima of this action functional (see Harland \& Popov \cite{Harland:2010ix} for the case of $\sSpin(7)$-instantons): in terms of our present setting of contact instantons on $K$-contact manifolds $M$ with gauge group $\sSU(r)$, the action functional is
\begin{equation}\label{eq:action}
 S\ =\   -\tfrac12 \int_M\operatorname{tr}\big\{\CF\wedge\star_5\CF\mp\eta\wedge\CF\wedge\CF\big\}~.
\end{equation}
Upon variation with respect to $\CA$, we find $\partial_\CA\star_5\!\CF=\pm\dd\eta\wedge\CF$. It is then rather straightforward to show that\footnote{See also \cite{Kallen:2012cs}.}
\begin{equation}
\begin{aligned}
 \operatorname{tr}\big\{\big(\CF+\kappa\,{\star_5(\eta\wedge\CF)}\big)\wedge\star_5 \big(\CF+\kappa\,{\star_5(\eta\wedge\CF)}\big)  \big\}+ \kappa^2 \operatorname{tr}\big\{\xi\lrcorner\CF\wedge\star_5 \left(\xi\lrcorner\CF\right) \big\} \ = \\
 =\ (1+\kappa^2)\operatorname{tr}\left\{\CF\wedge\star_5\CF+\frac{2\kappa}{1+\kappa^2}\,\eta\wedge\CF\wedge\CF\right\},\kern1cm
 \end{aligned}
\end{equation}
where $\kappa$ is some constant. Here, we made use of the contact metricity \eqref{eq:CMS}. Hence, $\kappa^2=1$ and
\begin{equation}
\begin{aligned}
S\ =\ -\tfrac14\int_M
  \operatorname{tr}\big\{\big(\CF\mp{\star_5(\eta\wedge\CF)}\big)\wedge\star_5 \big(\CF\mp{\star_5(\eta\wedge\CF)}\big) +\xi\lrcorner\CF\wedge\star_5 \left(\xi\lrcorner\CF\right) \big\}~.
  \end{aligned}
\end{equation}
Therefore, the absolute minima of the action functional $S$ (i.e.~$S=0$) are obtained whenever $\CF=\pm{\star_5(\eta\wedge\CF)}$ and $\xi\lrcorner\CF=0$. However, the last equation is implied by the former thus, $\CF=\pm{\star_5(\eta\wedge\CF)}$ is the sole equation. In addition, we have a bound on the Yang--Mills action functional
\begin{equation}
  -\!\tfrac12\int_M\operatorname{tr}\big\{\CF\wedge\star_5\CF\}\ \geq\ 
   \mp\tfrac12\int_M\operatorname{tr}\big\{\eta\wedge\CF\wedge\CF\big\}~.
\end{equation}
where the equality is achieved on solutions to $\CF=\pm{\star_5(\eta\wedge\CF)}$. 

Finally, we would like to emphasise that the equation  $\partial_\CA\star_5\!\CF=\pm\dd\eta\wedge\CF$ together with its action functional \eqref{eq:action} naturally appear when dimensionally reducing to five dimensions (anti-)self-dual three-form fields on six-dimensional manifolds that arise as circle fibrations over five-dimensional manifolds (including their non-Abelianisation from a five-dimensioanl point of view) \cite{Witten:2009at,Linander:2011jy}.
}

\paragraph{Higher-dimensional extension.}
The above considerations can be extended to $(2,m)$-paraconformal $K$-contact manifolds. For the sake of concreteness, let us focus on the case when $H_\FC\cong S\otimes\odot^q\tilde S$ for two rank-$2$ bundles $S$ and $\tilde S$ and $q+1=m$. Take $Z:=\PP(\tilde S^\vee)$ as before and endow it with the $F$-structure generated by
\begin{equation}
\begin{aligned}
 V_{\alpha}\ :=\ \pi^{\dot\alpha_1}\cdots \pi^{\dot\alpha_q} \left(E_{\alpha\dot\alpha_1\cdots\dot\alpha_q} +E_{\alpha\dot\alpha_1\cdots\dot\alpha_q}\lrcorner\omega^H{}_{\dot\beta}{}^{\dot\gamma}\,\pi_{\dot\gamma}\,\der{\pi_{\dot\beta}}\right),\qquad
 V^{\dot\alpha}\ :=\ \der{\hat\pi_{\dot\alpha}}~,\\
 \\[-10pt]
  \xi'\ :=\ \xi +\xi\lrcorner\omega^H{}_{\dot\alpha}{}^{\dot\beta}\left(\pi_{\dot\beta}\,\der{\pi_{\dot\alpha}}+\hat\pi_{\dot\beta}\,\der{\hat\pi_{\dot\alpha}}\right).\kern2.5cm
  \end{aligned}
 \end{equation}
Then upon following the same steps given in the proofs of Theorem \ref{thm:CRS-PWtrafo} and Theorem \ref{thm:FS-PWtrafo}, respectively, it is rather easy to see that $F$-flat rank-$r$ complex vector bundles $E_Z\to Z$ that are holomorphically trivial up the fibres of $\pi:Z\to M$ are in one-to-one correspondence with rank-$r$ complex vector bundles $E_M\to M$ that come with a connection $\partial_\CA$ such that the curvature $\CF=\partial_\CA^2$ obeys
\begin{equation}\label{eq:ContactWard}
\begin{aligned}
 \CF_{\alpha(\dot\alpha_1\cdots\dot\alpha_q\,\beta\dot\beta_1\cdots\dot\beta_q)}\ =\ E_{\alpha(\dot\alpha_1\cdots\dot\alpha_q}\CA_{\beta\dot\beta_1\cdots\dot\beta_q)}-E_{\beta(\dot\beta_1\cdots\dot\beta_q}\CA_{\alpha\dot\alpha_1\cdots\dot\alpha_q)}\, +\kern1.5cm\\
 +\, [\CA_{\alpha(\dot\alpha_1\cdots\dot\alpha_q},\CA_{\beta\dot\beta_1\cdots\dot\beta_q)}]-f_{\alpha(\dot\alpha_1\cdots\dot\alpha_q\,\beta\dot\beta_1\cdots\dot\beta_q)}{}^{\gamma\dot\gamma_1\cdots\dot\gamma_q}\CA_{\gamma\dot\gamma_1\cdots\dot\gamma_q}\ &=\ 0~,\\
 \CF_{\alpha\dot\alpha_1\cdots\dot\alpha_q\,\xi}\ =\ E_{\alpha\dot\alpha_1\cdots\dot\alpha_q}\CA_{\xi}-\xi\CA_{\alpha\dot\alpha_1\cdots\dot\alpha_q}\, +\kern1.5cm\\
 +\, [\CA_{\alpha\dot\alpha_1\cdots\dot\alpha_q},\CA_{\xi}]-f_{\alpha\dot\alpha_1\cdots\dot\alpha_q\,\xi}{}^{\gamma\dot\gamma_1\cdots\dot\gamma_q}\CA_{\gamma\dot\gamma_1\cdots\dot\gamma_q}\ &=\ 0~.
 \end{aligned}
\end{equation}
The equation $ \CF_{\alpha(\dot\alpha_1\cdots\dot\alpha_q\,\beta\dot\beta_1\cdots\dot\beta_q)}=0$ resembles the equation of the self-dual model represented by the $B_q$-series in Ward's classification scheme of higher-dimensional completely solvable gauge-field equations  \cite{Ward:1983zm}.\footnote{As shown in \cite{Wolf:2009ep}, Ward's $B_q$-series appears to be intimately connected with integrable superstring sigma-models.} Therefore, we may regard \eqref{eq:ContactWard} as the contact version of Ward's $B_q$-series. Notice that one may obtain contact versions of the other models given in \cite{Ward:1983zm} by using other paraconformal structures on the contact distribution.

\section{Supersymmetric extensions}\label{sec:SUSY}

Finally, we would like to extend the contact instanton equation supersymmetrically. To this end, we need supermanifolds and we shall work with supermanifolds in the sense of Manin \cite{Manin:1988ds}. In particular, we call a ringed space $(M,\CE_M)$ a real supermanifold of dimension $m|n$ provided $M$ is a topological space and $\CE_M$ is a sheaf of supercommutative rings on $M$ such that  the body $M_0:=(M,\CE_0:=\CE_M/\CI)$ is a smooth manifold of dimension $m$, where $\CI$ is the ideal subsheaf in $\CE_M$ that consists of all nilpotent elements and, in addition, we have $\CE_M\cong\CE_0(\Lambda^\bullet\FR^n)$.\footnote{Due to Batchelor \cite{Batchelor:1979}, any smooth supermanifold has $\CE_M\cong\CE_0(\Lambda^\bullet\FR^n)$. This is not true, however, in the complex category.} For supersymmetric extensions in the four-dimensional case, such as supersymmetric extensions of Penrose' non-linear graviton and Ward's construction, see e.g.~\cite{Manin:1988ds,Merkulov:1991kt,Merkulov:1992,Merkulov:1992qa,Merkulov:1992b,Witten:2003nn,Popov:2004rb,Berkovits:2004jj,Boels:2006ir,Wolf:2007tx,Mason:2007ct}. For brevity, we shall write $M$ instead of $(M,\CE_M)$.

\paragraph{Contact supermanifolds.}
For the moment, let us assume that $M$ is a real supermanifold of dimension $2m+1|2n$. In the next paragraph, we restrict ourselves to the case when $(m,n)=(2,\CN)$ with $\CN\in2\NN$. An even (bosonic) contact structure on $M$ is a rank-$2m|2n$ distribution $H\hookrightarrow TM$ in the tangent bundle that is maximally non-degenerate in the sense that the Frobenius form 
\begin{equation}  
 \Phi\,:\, H\wedge  H\ \to\ L\ =\ TM/ H~,\ewith \Phi(X,Y)\ :=\ [X,Y\} \mbox{ mod }  H
\end{equation}
is non-degenerate on $H$ for any $X,Y\in\Gamma(M,TM)$. Here, `$[\cdot,\cdot\}$' denotes the supercommutator.   As before, $H$ can be defined dually as the kernel of a nowhere vanishing even (bosonic) differential one-form $\eta$, that is, $H=\ker \eta$. On the body $M_0$ of $M$, the non-degeneracy requirement of $H$ is again equivalent to saying that $\eta\wedge(\dd\eta)^m\neq0$. However, on $M$ this will no longer be true since $(\dd\theta)^k\neq 0$ for any $k\in\NN$ and any odd (fermionic) coordinate $\theta$.  If $M$ is equipped with a non-degenerate even contact structure, then we call it a contact supermanifold. 

In addition, we may define a contact metric structure as in the purely even case, that is, the datum $(\eta,\xi,g,\phi)$ on $M$ is called a contact metric structure on a contact supermanifold if $g$ is a supermetric and $\phi$ is an even endomorphism of $TM$ such that the equations \eqref{eq:CMS} are satisfied. We then call $M$ a contact metric supermanifold. Likewise, we call $M$ a $K$-contact supermanifold, provided the Reeb vector field is Killing with respect to the supermetric $g$ and the Levi-Civita connection.

Following our previous discussion, we now introduce paraconformal contact supermanifolds. In particular, if let $M$ be an $(2m+1|\CN m)$-dimensional contact supermanifold with contact distribution $H$, then we would like to introduce a $(2|\CN,m|0)$-paraconformal structure according to $\sigma:H_\FC\overset{\cong}{\to}S\otimes\tilde S$, where now $S$ is a rank-$2|\CN$ complex vector bundle while $\tilde S$ is of rank $m|0$. As before, we shall equip $S$ and $\tilde S$ with real structures $\tau$ and $\tilde\tau$ which have the same properties as in the purely even setting, so $m,\CN\in2\NN$. If, in addition, we also have  $(\eta,\xi,g,\phi)$ such that $g$ and $\phi$ are compatible with the paraconformal structure (i.e.~$g|_H=\varepsilon\otimes\tilde\varepsilon$ and $\phi|_H=\phi^E\otimes\mbox{id}^{\tilde S}$ as in the purely even setting), we shall speak of paraconformal contact metric supermanifolds and of paraconformal $K$-contact supermanifolds provided $\xi$ is Killing (we shall drop the prefix ``para" when $m=2$). 

Finally, we point out that we have again a decomposition of the differential two-forms as in \eqref{eq:TwoForms}. This decomposition is needed in the next paragraph.

{\remark
We would like to emphasise that our paraconformal contact supermanifolds are similar to the chiral supermanifolds used in four dimensions to describe self-dual supergravity and its twistor theory, see e.g.~\cite{Wolf:2007tx,Mason:2007ct}. Our main motivation for this framework is to find a supersymmetric extension  of the contact instanton equation in spirit of self-dual supersymmetric Yang--Mills theory. However, if one has (full)  supergravity applications in mind, then one should introduce different structures than those considered here since one has to incorporate various torsion constraints, see e.g.~Howe \cite{Howe:1981gz}.  
}

\paragraph{Penrose--Ward transform and supersymmetric contact instantons.}
Now it is rather easy to generalise the contact instanton equation \eqref{eq:CIE-2} supersymmetrically. For the sake of concreteness we shall only consider this case but the supersymmetric extension of e.g.~\eqref{eq:ContactWard} can be constructed similarly.

Consider a $(5|2\CN)$-dimensional conformal $K$-contact supermanifold $M$ with  $\sigma:H_\FC\overset{\cong}{\to}S\otimes\tilde S$. We then may introduce frame fields $E_{A\dot\alpha}$ and $\xi$ with $A=(\alpha,i)$, where $\alpha,\dot\alpha,\ldots=1,2$ and $i,j,\ldots=1,\ldots,\CN$. So $E_{\alpha\dot\alpha}$ and $\xi$ constitute the even (bosonic) frame fields while $E_{i\dot\alpha}$ the odd (fermionic) ones. We define the Cauchy--Riemann supertwistor space as in the purely even setting \eqref{eq:TS}, that is, $Z:=\PP(\tilde S^\vee)$. We then introduce an almost $F$-structure given by
\begin{subequations}\label{eq:FS-super}
\begin{equation}
 F\ :=\ \langle V_A,\xi',V^{\dot\alpha}\rangle
 \end{equation} 
with
\begin{equation}
\begin{aligned}
 V_A\ :=\ \pi^{\dot\alpha}\underbrace{\left(E_{A\dot\alpha} +E_{A\dot\alpha}\lrcorner\omega^H{}_{\dot\beta}{}^{\dot\gamma}\,\pi_{\dot\gamma}\,\der{\pi_{\dot\beta}}\right)}_{=:\ V_{A\dot\alpha}},\qquad
 V^{\dot\alpha}\ :=\ \der{\hat\pi_{\dot\alpha}}~,\\
   \xi'\ :=\ \underbrace{\xi +\xi\lrcorner\omega^H{}_{\dot\alpha}{}^{\dot\beta}\left(\pi_{\dot\beta}\,\der{\pi_{\dot\alpha}}\right.}_{=:\ V_\xi}\left.+\hat\pi_{\dot\beta}\,\der{\hat\pi_{\dot\alpha}}\right).\kern1.5cm
\end{aligned}
 \end{equation}
 \end{subequations}
which is very similar as the one given in the purely even setting. Then we have the following result:

\vspace{10pt}
{\theorem
Let $M$ be a $(5|2\CN)$-dimensional conformal $K$-contact supermanifold with Levi-Civita connection $\nabla$ and curvature $\CR$. Consider the restrictions of $\CR$ to $\Omega_-^2(M)$ and $\eta\wedge\Omega^1(M)$, respectively, that map into $\Omega_-^2(M)$, i.e.~$\CR_-^-:\Omega_-^2(M)\to \Omega_-^2(M)$ and $\CR_\eta^-:\eta\wedge\Omega^1(M)\to \Omega_-^2(M)$. The almost $F$-structure \eqref{eq:FS-super} on the Cauchy--Riemann supertwistor space of $M$ is integrable if and only if the totally trace-free parts of $\CR_-^-$ and $\CR_\eta^-$ vanish.
} 
 
\vspace{15pt}
\noindent
The proof of this result is similar to the one given for Theorem \ref{thm:FS}. We therefore refrain from repeating the steps here but would like to refer to references \cite{Wolf:2007tx,Mason:2007ct}, where details on the construction in the four-dimensional setting can be found including the decomposition of the curvature in irreducible pieces. Notice that special care must be taken when $\CN=4$.
 
Now we have all the ingredients to state the supersymmetric extension of Theorem \ref{thm:FS-PWtrafo}:

\vspace{10pt}
{\theorem\label{thm:FS-PWtrafo-super}
Let $M$ be a $(5|2\CN)$-dimensional conformal $K$-contact supermanifold with Cauchy--Riemann supertwistor space $\pi:Z\to M$ and (integrable) $F$-structure \eqref{eq:FS-super}. Then there is a one-to-one correspondence between
\begin{itemize}\setlength{\itemsep}{-1mm}
\item[{(i)}] $F$-flat rank-$r|s$ complex supervector bundles $E_Z\to Z$ such that $E_Z$ is holomorphically trivial up the fibres $\pi:Z\to M$ and
\item[{(ii)}] rank-$r|s$ complex supervector bundles $E_M\to M$ equipped with a connection $\partial_{\CA}$ and curvature $\CF=\partial_{\CA}^2$ such that $\CF\in\Omega^2_+(M,\operatorname{End}E_M)$, that is, $\CF_-=0=\CF_\eta$.
\end{itemize}  
}

\vspace{15pt}
\noindent
To prove this assertion, we can follow the arguments given when proving Theorem \ref{thm:CRS-PWtrafo} and Theorem \ref{thm:FS-PWtrafo}. Eventually, we find the equations
\begin{equation}\label{eq:SuperContactInst}
\begin{aligned}
 \CF_{A(\dot\alpha\,B\dot\beta)}\ :=\ E_{A(\dot\alpha}\CA_{B\dot\beta)}-E_{B(\dot\beta}\CA_{A\dot\alpha)}+[\CA_{A(\dot\alpha},\CA_{B\dot\beta)}\}-f_{A(\dot\alpha\,B\dot\beta)}{}^{C\dot\gamma}\CA_{C\dot\gamma}\ &=\ 0~,\\
 \CF_{A\dot\alpha\,\xi}\ :=\ E_{A\dot\alpha}\CA_{\xi}-\xi\CA_{A\dot\alpha}+[\CA_{A\dot\alpha},\CA_{\xi}\}-f_{A\dot\alpha\,\xi}{}^{C\dot\gamma}\CA_{C\dot\gamma}\ &=\ 0~,
 \end{aligned}
\end{equation}
for the superfields  $\CA_{A\dot\alpha}$ and $\CA_\xi$. These equations are the compatibility conditions of the auxiliary linear problem
\begin{subequations}
 \begin{equation}
  \lambda^{\dot\alpha}(V_{A\dot\alpha}+\CA_{A\dot\alpha})\psi\ =\ 0\eand (V_\xi+\CA_\xi)\psi\ =\ 0~,
\end{equation}
where $(\lambda_{\dot\alpha}):=(1,\lambda)$ and 
\begin{equation}
 V_{A\dot\alpha}\ =\ E_{A\dot\alpha} -E_{A\dot\alpha}\lrcorner\omega^H{}_{\dot\beta}{}^{\dot\gamma}\,\lambda_{\dot\gamma}\lambda^{\dot\beta}\,\der{\lambda}\eand
  V_\xi\ =\ \xi - \xi\lrcorner\omega^H{}_{\dot\alpha}{}^{\dot\beta}\,\lambda_{\dot\beta}\lambda^{\dot\alpha}\,\der{\lambda}~.
\end{equation}
\end{subequations}
The system \eqref{eq:SuperContactInst} can be understood as a five-dimensional extension of the constraint system of self-dual supersymmetric Yang--Mills theory.

\vspace{10pt}
\paragraph{Acknowledgements.}
I am very grateful to M.~Dunajski, D.~Harland, M.~Henningson, L.~Mason, and C.~S{\"a}mann for important discussions, questions and suggestions.

\end{document}